\documentclass[aps,pre,showpacs,floatfix,groupedaddress,twocolumn]{revtex4-1}
\usepackage{bm,color,amsmath,amssymb,siunitx,blindtext}
\usepackage{graphicx,epstopdf}

\usepackage[latin1]{inputenc}
\newcommand{\beq}{\begin{equation}}
\newcommand{\eeq}{\end{equation}}
\newcommand{\bea}{\begin{eqnarray}}
\newcommand{\eea}{\end{eqnarray}}

\newcommand{\cev}[1]{\reflectbox{\ensuremath{\vec{\reflectbox{\ensuremath{#1}}}}}}

\begin{document} \title{Mapping of uncertainty relations between continuous and discrete time.}  
\author{Davide Chiuchi\`{u}} \affiliation{Biological Complexity Unit, Okinawa Institute of Science and Technology and Graduate University, Onna, Okinawa 904-0495.} 
 \author{ Simone Pigolotti} \email{simone.pigolotti@oist.jp} 
\affiliation{Biological Complexity Unit, Okinawa Institute of Science and Technology and Graduate University, Onna, Okinawa 904-0495.}  

\begin{abstract} Lower bounds on fluctuations of thermodynamic currents depend on the nature of time: discrete or continuous. To understand the physical reason, we compare current fluctuations in discrete-time Markov chains and continuous-time master equations. We prove that current fluctuations in the master equations are always more likely, due to random timings of transitions.  This comparison leads to a mapping of the moments of a current between discrete and continuous time. We exploit this mapping to obtain uncertainty bounds.  Our results reduce the quests for uncertainty bounds in discrete and continuous time to a single problem. 
 \end{abstract}


\maketitle
Fluctuations play an important role in the thermodynamics of small-scale systems. Stochastic thermodynamics studies how these fluctuations affects observables such as the heat exchanged between a system and its environment,  the work output of a small device, and the device's efficiency \cite{KenSekimotoEnergetics,Seifert2012,PhysRevLett.78.2690,PhysRevE.60.2721,PhysRevE.90.052145}. These observables can be generally expressed in terms of stochastic currents.

Although most properties of stochastic currents are system dependent, some general results, known as uncertainty relations, have been derived in recent years \cite{PhysRevLett.114.158101,gingrich2016dissipation,polettini2016tightening,horowitz2017proof,pigolotti2017generic,dechant2017current,PhysRevE.93.052145,Pietzonka2016}. In general, uncertainty relations provide bounds on the fluctuations of stochastic currents. A main result, first observed in \cite{PhysRevLett.114.158101} and rigorously proven in \cite{gingrich2016dissipation}, states that the large-deviation function of a generic current, at steady state, is broader than predicted by linear response theory. A direct consequence is a bound on the variance of a current in terms of its mean and the entropy production rate.  This result was derived for master equations in the long-time limit \cite{gingrich2016dissipation} and then for finite time \cite{horowitz2017proof}. The same bound holds for continuous state-space  Langevin equations \cite{pigolotti2017generic,dechant2017current}. These results suggested that the uncertainty bound should be general and rather insensitive to the details of the system. 

It therefore came as a surprise when it was reported that the uncertainty bound \cite{PhysRevLett.114.158101, gingrich2016dissipation} does not hold for a system described by a discrete-time Markov chain \cite{shiraishi2017finite}. A more recent paper \cite{proesmans2017discrete}, following the mathematical strategy of \cite{gingrich2016dissipation}, proved a looser bound on the rate function for Markov chains. Besides their theoretical interest, discrete-time bounds have a practical relevance since periodically driven small-scale systems  \cite{rosas2017stochastic}  can be thought of as discrete-time processes. Despite these results, it remains counter-intuitive why the stationary statistics of currents should depend on whether time is discrete or not.  

In this paper, we systematically compare current fluctuations in continuous and discrete time.  By associating a Markov chain with each master equation, we show that the variance of a generic current in the master equation equals that in the corresponding Markov chain plus a non-negative correction term. This difference originates from fluctuations in the total number of transitions, as previously observed for the diffusion coefficient \cite{derrida1983velocity,maes1988discrete,Koza1999}. We generalize this result to arbitrary systems, arbitrary currents, and higher cumulants. We further demonstrate that the current large-deviation function is broader for discrete time than for continuous time.  The expression of the correction term for the variance establishes a rule to export bounds derived for continuous-time processes to discrete ones and vice versa. In particular, the bound in \cite{proesmans2017discrete} leads to a bound for the continuous case [Eq. \eqref{eq:vdbcont}] which is tighter than that  in \cite{gingrich2016dissipation}.

\begin{figure}[tb]
\centering 
\includegraphics[width=\linewidth]{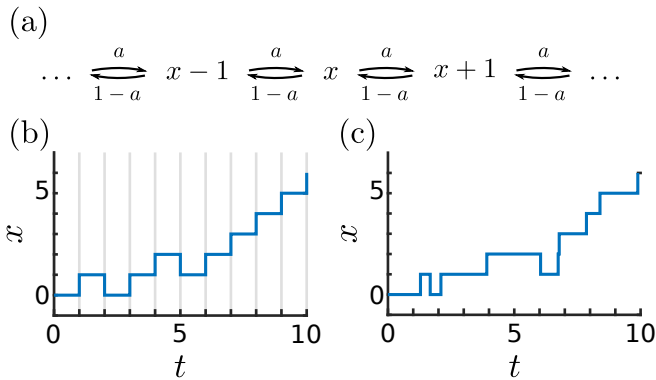}
\caption{Biased random walk. (a) Transition network. (b) Sample trajectory for the Markov chain \eqref{eq:RW_markov}. (c) Sample trajectory for the master equation \eqref{eq:RW_master}. Notice the fluctuations in the jump times in (c), which are absent in (b).  \label{fig:RW_example}}
\end{figure}

{\em Biased random walk.} We introduce our idea with the example of a biased random walk [see Fig. \ref{fig:RW_example}a]. We compare two different models. In the first one, time is discrete and the probability distribution is governed by the Markov chain
\begin{equation}\label{eq:RW_markov}
P_{x}(t+1)=a P_{x-1}(t) + (1-a) P_{x+1}(t),
\end{equation}
where $P_x(t)$ is the probability that the system is in position $x$ at time $t$ and $0\le a \le 1$ is a parameter determining the bias.
In the second model, time is continuous and the system evolves according to
the master equation
\begin{equation}\label{eq:RW_master}
\frac{\mathrm{d} P_{x}(t)}{\mathrm{d}t}=a P_{x-1}(t) + (1-a) P_{x+1}(t)-P_{x}(t).
\end{equation}
In both cases, we consider the empirical integrated current 
\begin{equation}\label{eq:RW_current}
J(t)=n^+(t)-n^-(t)
\end{equation}
where $n^+(t)$ and $n^-(t)$ are the total numbers of transitions where $x$ increases or decreases, respectively, up to a time $t$. 
Let us look at the moments of $J(t)$ in the two cases. For the discrete-time model of Eq. \eqref{eq:RW_markov}, it is known that 
\begin{eqnarray}\label{moments:d}
\langle J \rangle_d &=& (2a-1)t\nonumber\\  
 \sigma^2_{J,d} &=& 4a(1-a)  t
\end{eqnarray}
where $\sigma^2_{J,d}=\langle J^2\rangle_d-\langle J\rangle_d^2$. From now on we use the notation $\langle \dots \rangle_d$ and $\langle \dots \rangle_c$ for averages over the discrete-time and continuous-time processes, respectively.
The same quantities for the model of Eq. (\ref{eq:RW_master}) read 
\begin{eqnarray}\label{moments:c}
\langle J\rangle_c&=& (2a-1)t \nonumber\\
\sigma^2_{J,c}&=&t .
\end{eqnarray}
Note that the average current is equal in the two cases, whereas the variance is larger or equal in the continuous case. 
In particular, we have
 \begin{equation}\label{rw_sigma}
\sigma^2_{J,c}-\sigma^2_{J,d}=t(2a-1)^2=\frac{\langle J \rangle_d^2}{t}=\frac{\langle J \rangle_c^2}{t} .
\end{equation}

Note that the difference can also be written as $\sigma^2_{J,c}-\sigma^2_{J,d}=(2a-1)^2(\langle n^2\rangle_c-\langle n\rangle_c^2)$, i.e., the enhanced fluctuations in the continuous-time current originate from the fluctuations in the total number of transitions observed in a given time interval [see the comparison between Fig. \ref{fig:RW_example}(a) and \ref{fig:RW_example}(b)]. This effect has been previously studied for general random walks \cite{derrida1983velocity,maes1988discrete,Koza1999}.  In the following, we will show that this result holds for general systems and  general currents. 

{\em General theory. } Let us consider a general system described by a Markov chain
\begin{equation}
P_k(t+\tau)=\sum\limits_{l=1}^M A_{kl} P_l(t)
\end{equation}
where $P_k(t)$ is the probability of being in state $k$ at time $t$, $\tau$ is the timestep of the process, and $0 \le A _{kl}\le 1$ are the transition probabilities from state $l$ to $k$, with $1\le k,l\le M$. Note that self-transitions are included through the diagonal terms $A_{ll}$. Conservation of probability requires $\sum_k A_{kl}=1$ $\forall l$.
In parallel, we consider the master equation
\begin{equation}
\frac{\mathrm{d} P_k(t)}{\mathrm{d} t}=\sum_{l=1}^M W_{kl}P_{l}(t)
\end{equation}
with transition rates from state $l$ to state $k$ given by $W_{kl}\ge 0$ for $k\neq l$. Conservation of probability here requires that $W_{ll}=-\sum_{k;k\neq l}W_{kl}$ $\forall l$.
 From now on, we assume ergodicity and that $A_{kl}>0$ if and only if $A_{lk}>0$ for all $k\neq l$, and similarly for $W_{kl}$.

To link a given Markov chain and a given master equation, we introduce the mapping
\begin{equation}\label{eq:mapping}
\hat{A}=\hat{\mathbb{I}}+\tau\hat{W}
\end{equation}
where $\hat{A}$ is  the matrix having elements $A_{kl}$, $\hat{W}$ is the matrix having elements $W_{kl}$, and
$\hat{\mathbb{I}}$ is the identity matrix. Notice that, for any matrix $\hat{A}$ defining a Markov chain and any timestep $\tau$, the mapping in Eq. \eqref{eq:mapping} yields a unique, well-defined master equation. Conversely, when using Eq. \eqref{eq:mapping} to map a master equation into a Markov chain, $\tau$ is a free parameter. However, $\tau$ should be chosen such that
\begin{equation}\label{eq:tau_lim}
\tau\leq \frac{1}{\max\limits_{l} \left[-W_{ll}\right]}.
\end{equation}
to ensure that all the diagonal terms $A_{ll}$ are non-negative. Physically, this condition means that the timestep of the associated Markov chain should be small enough to resolve all the fast time scales of the process.

We now consider two  processes linked by Eq.\eqref{eq:mapping} and  study, for both of them, a generalized empirical current
\begin{equation}
j(t)=\frac{1}{t}\sum_{k,l}~j_{kl}~n_{kl}(t)
\end{equation}
where $j_{kl}$ is a given antisymmetric real matrix and $n_{kl}(t)$ is the number of transitions $l\rightarrow k$ observed up to a time $t$.
To compute the moments of $j$ at large times, we consider its scaled cumulant generating function. In the discrete case, it reads (see \cite{ellis2007entropy,TOUCHETTE20091} and the Appendix):
\begin{equation}\label{psid}
\psi_d(q)=\lim_{t\rightarrow \infty}\frac{1}{t}\ln \left\langle e^{q t j(t)} \right \rangle_d = \frac{\ln  ~\lambda(q)}{\tau}
\end{equation}
where $\lambda(q)$ is the dominant eigenvalue of the tilted matrix $\hat{B}$ with components
\begin{eqnarray}
B_{kl}&=&A_{kl} ~ e^{q j_{kl}}=\tau W_{kl} ~ e^{q j_{kl}}\quad k\neq l \nonumber\\
B_{ll}&=&A_{ll} =1+\tau W_{ll} \label{eq:tilted_matrix}.
\end{eqnarray}
The Perron-Frobenius theorem ensures that $\lambda(q)$ is real, positive, and non degenerate for all real values of $q$. Similarly, it can be shown (see \cite{Koza1999,PhysRevE.92.042132,Lebowitz1999,Budini2014} and the Appendix) that  the scaled cumulant generating function in the continuous case reads
\begin{equation}\label{psic}
\psi_c(q)=\lim_{t\rightarrow \infty}\frac{1}{t}\ln \left\langle e^{q t j(t)} \right \rangle_c = \frac{\lambda(q)-1}{\tau} .
\end{equation}

The scaled moments of $j$ can be computed from $\psi_d(q)$ and $\psi_c(q)$. 
For the averages, we obtain
\begin{eqnarray}\label{eq:mean}
\langle j\rangle_d &=&\psi'_d(0)=\frac{\lambda'(0)}{\tau}\nonumber\\
\langle j\rangle_c&=&\psi'_c(0)=\frac{\lambda'(0)}{\tau}
\end{eqnarray}
where primes denote derivatives respect to $q$ and we used $\lambda(0)=1$. The average generalized currents are therefore equal in discrete and continuous time. Instead, the scaled variances are
\begin{eqnarray}\label{eq:var}
\tilde{\sigma}^2_{j,d}&=&  \lim_{t\rightarrow\infty} t ~\sigma ^2_{j,d}=\psi''_d(0) =\frac{\lambda''(0)-[\lambda'(0)]^2}{\tau} \nonumber\\
\tilde{\sigma}^2_{j,c}&=&  \lim_{t\rightarrow\infty} t ~\sigma ^2_{j,c}=\psi''_c(0) =\frac{\lambda''(0)}{\tau} .
\end{eqnarray}
Substituting Eq. \eqref{eq:mean} into Eq. \eqref{eq:var}, we find that
\begin{equation}\label{eq:var2}
\tilde{\sigma}^2_{j,c}=\tilde{\sigma}^2_{j,d}+\langle j\rangle^2 \tau .
\end{equation}
This result generalizes Eq.\eqref{rw_sigma}, to an arbitrary current in an arbitrary system.
The same procedure can be carried out to explicitly compute differences of higher cumulants between the discrete and the continuous case. 

\begin{figure}
\includegraphics[width=\linewidth]{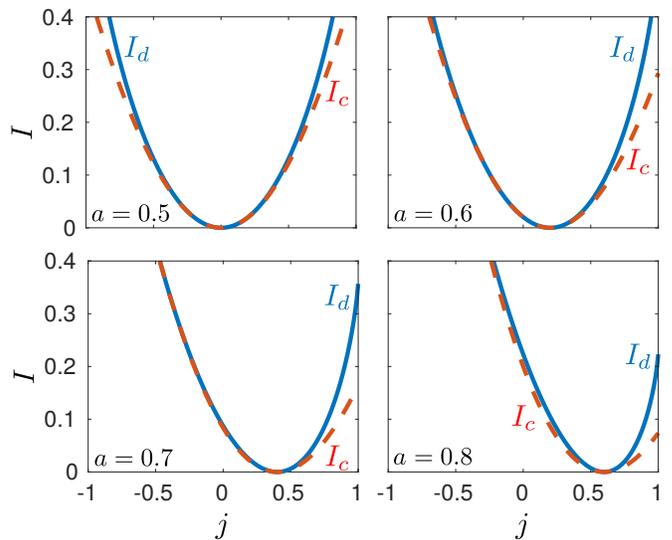}
\caption{Comparison of the rate functions for the discrete and continuous cases [see Eqs. \eqref{eq:ldt}] for the biased random walk. The four panels correspond to different choices of the bias $a$, as shown in the legend. The rate functions are computed analytically for a  three-state random walk with periodic boundary conditions and $\tau=1$. Note that the comparison is possible only in the interval shown in the figure, since $j\in[-1,1]$ in the discrete case.  \label{fig:rate_functions}}
\end{figure}

In general, $j(t)$ satisfies a large deviation principle \cite{ellis2007entropy,TOUCHETTE20091,Maes2008},
$j(t)\sim e^{-t I(j)}$,
where the discrete and continuous rate functions $I=I_d(j)$ and $I=I_c(j)$, respectively, are given by the G\"{a}rtner-Ellis theorem \cite{TOUCHETTE20091}
\begin{eqnarray}\label{eq:gartner}
I_d(j)&=&\sup_{q\in \Re}~[ qj-\psi_d(q)]\nonumber\\
I_c(j)&=&\sup_{q\in \Re}~[ qj-\psi_c(q)] .
\end{eqnarray}
From Eqs. (\ref{psid}) and (\ref{psic}) one has $\psi_d(q)\le\psi_c(q)$ for all $q\in\Re$. We therefore conclude from Eq. \eqref{eq:gartner} that 
\begin{equation} \label{eq:ldt}
I_d(j)\ge I_c(j) .
\end{equation}
Equation \eqref{eq:ldt}  is one of the main results of this paper. It states that large current fluctuations are always less likely in discrete time than in continuous time. A comparison of the discrete and continuous rate functions is presented in Fig. \ref{fig:rate_functions} for the biased random walk at different values of the bias. Interestingly, the two rate functions are different also at equilibrium, i.e., when  $\langle j\rangle = 0$, as illustrated for the unbiased case $a=0.5$ of  Fig. \ref{fig:rate_functions}.  Note that Eqs. \eqref{rw_sigma} and \eqref{eq:var2} predict $\sigma^2_{j,d}=\sigma^2_{j,c}$ in this case. However, the two rate functions are identical only if approximated by low-order polynomials, as the differences in cumulants of order $4$ and above are not proportional to $\langle j\rangle$. 

{\em Uncertainty bounds.}  It has been recently observed that discrete-time processes satisfy looser uncertainty bounds than continuous-time processes \cite{shiraishi2017finite, proesmans2017discrete}. The relations derived in the preceding section yield an explicit mapping for the moments of generalized currents. They can therefore be used to systematically transform bounds for continuous-time processes \cite{PhysRevLett.114.158101,gingrich2016dissipation,PhysRevE.93.052145} to bounds for discrete-time processes and vice versa.

Most uncertainty relations bound the ratio between the variance and the squared average of a generalized current. Using Eqs. \eqref{eq:mean} and \eqref{eq:var2}, we obtain the mapping
\begin{equation}\label{fanomapping}
\frac{\tilde{\sigma}^2_{j,c}}{\langle j \rangle_c^2}=\frac{\tilde{\sigma}^2_{j,d}}{\langle j \rangle_d^2}+\tau .
\end{equation}
We now investigate the consequences of Eq. \eqref{fanomapping}. 
For example, the Barato-Gingrich (BG) bound \cite{PhysRevLett.114.158101,gingrich2016dissipation,PhysRevE.93.052145} for continuous-time systems reads
\begin{equation}\label{eq:gingrich}
\frac{\tilde{\sigma}^2_{j,c}}{\langle j \rangle_c^2}\ge \frac{2}{\Sigma_c}
\end{equation}
where we introduced the dimensionless average entropy production rate
\begin{equation}\label{eq:ent}
\Sigma_c=\left\langle\frac{1}{t}\sum_{k\neq l} n_{kl}(t) \ln\left[\frac{W_{kl}P^{(st)}_l}{W_{lk}P^{(st)}_k}\right]\right\rangle_c
\end{equation}
and $P^{(st)}_l$ are the stationary probabilities, which are invariant under the mapping. Since $W_{lk}=\tau A_{lk}$ for $l\neq k$ and $\Sigma_c$ is the average of a generalized current, it immediately follows from eq. \eqref{eq:mean} that $\Sigma_d=\Sigma_c$. Using this property and substituting Eq. \eqref{fanomapping} into Eq. \eqref{eq:gingrich} we obtain a mapped BG bound for discrete processes
\begin{equation}\label{eq:gingrichmapped}
\frac{\tilde{\sigma}^2_{j,d}}{\langle j \rangle_d^2}\ge \frac{2}{\Sigma_d}-\tau.
\end{equation}
Similarly, we consider the Proesmans-Van den Broeck (PV) bound \cite{proesmans2017discrete} on discrete-time processes
\begin{equation}\label{eq:vdb}
\frac{\tilde{\sigma}^2_{j,d}}{\langle j \rangle_d^2}\ge\frac{2\tau}{e^{\Sigma_d\, \tau}-1} .
\end{equation}
With the same idea, we obtain a mapped PV bound on continuous processes
\begin{equation}\label{eq:vdbcont}
\frac{\tilde{\sigma}^2_{j,c}}{\langle j \rangle_c^2}\ge\left(\frac{2}{e^{\Sigma_c\, \tau}-1}+1 \right)\tau
\end{equation}
which holds for any choice of $\tau$ satisfying Eq.\eqref{eq:tau_lim}.

\begin{figure}[tb]
\centering 
\includegraphics[width=\linewidth]{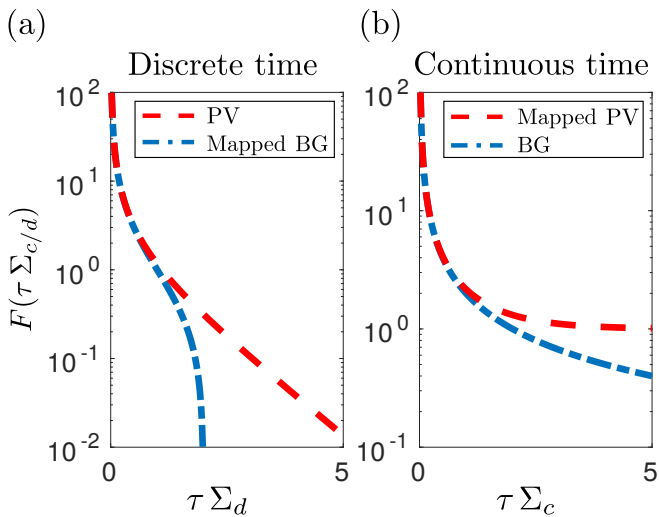}
\caption{Uncertainty bounds in the scaling form of Eq. \eqref{eq:scaling}. (a) Discrete-time bounds. The blue dot-dashed line and the red dashed lines represent the mapped Barato-Gingrich 
bound \eqref{eq:gingrichmapped} and  the Proesmans-Van den Broeck bound \eqref{eq:vdb} respectively. (b) Continuous-time bounds.  The blue dot-dashed line and the red dashed lines are the Barato-Gingrich bound \eqref{eq:gingrich}  and the mapped Proesmans-Van den Broeck bound \eqref{eq:vdbcont}, respectively.\label{fig:bound}}
\end{figure}

Note that the bounds of Eqs. \eqref{eq:gingrich} and \eqref{eq:gingrichmapped}-\eqref{eq:vdbcont} can all be cast in the scaling form
\begin{equation}\label{eq:scaling}
\frac{\tilde{\sigma}^2_{j,c/d}}{\langle j \rangle_{c/d}^2}\ge \tau F(\tau\, \Sigma_{c/d}) .
\end{equation}
The function $F(\tau\, \Sigma_{c/d})$ is represented in  Figs. \ref{fig:bound}(a) and \ref{fig:bound}(b) for the discrete and continuous bounds, respectively.
In the discrete case, the mapped BG bound is looser than the PV bound for all values of $\Sigma_d$ and becomes trivial for $\tau\,\Sigma_d\ge 2$. In the continuous case, the mapped PV bound is always tighter than the BG bound.  In particular, this bound does not tend to zero for large $\Sigma_c$. Indeed, a consequence of Eq. \eqref{eq:vdbcont} is
\begin{equation}\label{otherbound}
\frac{\tilde{\sigma}^2_{j,c}}{\langle j \rangle_c^2}\ge \tau .
\end{equation}
Equation \eqref{otherbound} means that one cannot arbitrarily reduce $\tilde{\sigma}^2_{j,c}$ at the expense of entropy production. This is another consequence of the unavoidable fluctuations in the number of transitions in master equations. Notice that Eq.\eqref{otherbound} can be also obtained as a consequence of the exponential bound ( see \cite{PhysRevE.93.052145}).

\begin{figure}[htb]
\centering 
\includegraphics[width=\linewidth]{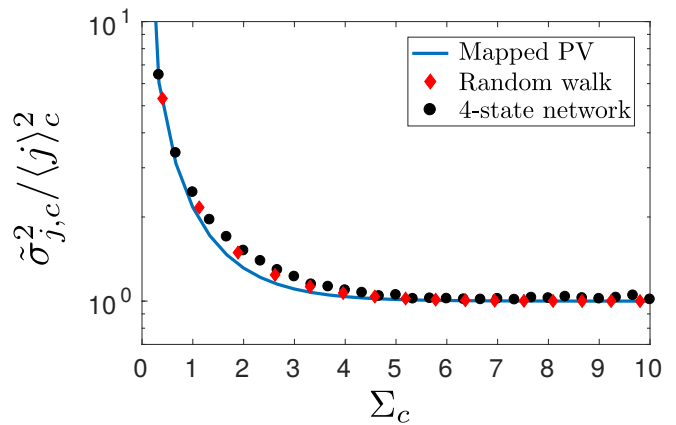}
\caption{Comparison of the bound \eqref{eq:vdbcont} with examples of a biased random walk and a fully connected four-state network.  In both cases, we set $\tau=1$. For the biased random walk, the current is $j=J/t$, where $J$ is given by Eq. \eqref{eq:RW_current} and entropy production is tuned by varying the bias parameter $a$. For the  four-state network, the parameters $j_{kl}$ defining the generalized current were independently drawn from a Gaussian distribution with zero average and unit variance. We then numerically minimize ${\tilde{\sigma}^2_{j,c}}/{\langle j \rangle_c^2}$ as a function of $\Sigma_c$, with the constraint $W_{ll}\geq-1$ for all $l$. The moments of the currents are computed with the method of Ref.  \cite{PhysRevE.92.042133}.   The minimization is carried out with the MATLAB patternsearch algorithm.  To avoid local minima, for each value of $\Sigma_c$, we perform $N=25$ different runs of patternsearch starting from different initial conditions. The smallest value among these realizations is plot in the figure. We verified that performing the minimization after a different random choice of the coefficients $j_{kl}$ leads to very similar results. 
Notice how the two systems are very close to the mapped Proesmans-Van den Broeck bound \eqref{eq:vdbcont} (blue solid line) for a broad range of $\Sigma_c$ values. \label{fig:4state}}
\end{figure}

To further corroborate our results, we compare the bound of Eq. \eqref{eq:vdbcont} with two different examples: the biased random walk and a fully connected four state network. In both cases, we set the value of $\tau$ yielding the tightest bound. For the four state network , we consider a random generalized current. For each value of $\Sigma_c$, we employ the method of Ref. \cite{PhysRevE.92.042133} and a constrained optimization algorithm to find the rates $W_{kl}$ that minimize the ratio $\tilde{\sigma}^2_{j,c}/\langle j \rangle_c^2$. Minimization is performed with the constraint $W_{ll}\geq-1$ $\forall l$. Results are shown in Fig. \ref{fig:4state} and suggest that the mapped PV bound can be saturated. Note that the asymptotic bound of Eq.\eqref{otherbound} can be tightened (see \cite{PhysRevE.93.052145}) as $\tilde{\sigma}^2_{j,c}/\langle j \rangle_c^2\ge -\mathcal{A}^{-1}$, where $\mathcal{A}=M^{-1}\sum_{l=1}^M W_{ll}$ is the mean activity. The numerical minimization in the figure approaches Eq. \eqref{otherbound}, thus suggesting that all the activities are approximately the same at the minimum of $\sigma_{j,c}^2/\langle j \rangle_c^2$.

{\em Conclusions.} In this paper, 
 we have shown that currents in master equations always present additional fluctuations due to random timings of transitions. Our work generalizes previous results on diffusion coefficients in discrete- and continuous-time random walks  \cite{derrida1983velocity,maes1988discrete,Koza1999} to arbitrary systems, arbitrary currents, and higher cumulants. Our theory predicts that the rate function of a current in a continuous-time system is always broader than its discrete counterpart.  We exploited this effect in Eqs. \eqref{eq:vdbcont} and \eqref{otherbound}. In particular,  Eq. \eqref{eq:vdbcont} is a lower bound on fluctuations of an arbitrary current that becomes significantly more stringent than  Eq.\eqref{eq:gingrich} for $\tau\,\Sigma_d\gg 1$. It can therefore be useful for highly dissipative systems, such as those found in biology  \cite{Nguyen13122016,Sartori2015}.

Our results are valid in the long-time limit. Generalization to finite time would require a study of sub dominant eigenvalues in the expressions of the scaled generating functions [Eqs. \eqref{psid} and \eqref{psic}]. Further, it would be interesting to consider more general mappings than Eq. \eqref{eq:mapping}. In this case, the tilted matrices for the discrete and continuous cases do not necessarily commute, so it is not trivial to find a relation between their leading eigenvalues.
Another problem is to assess whether the bound of Eq. \eqref{eq:vdbcont} is valid for Langevin equations or is particular to master equations. Such results would further clarify the role of continuous vs. discrete state space and time in determining current fluctuations.

\begin{acknowledgments}
We thank  A. Maritan, I. Neri, P. Pietzonka,  K. Proesmans, U. Seifert and C. Van den Broeck for comments on a preliminary version of this manuscript.
\end{acknowledgments}

\appendix
\section{Appendix}
In this appendix we demonstrate Eqs. \eqref{psid} and \eqref{psic}. Let us start with the discrete case. By definition
\begin{equation}\label{eq:psid_app}
\psi_d(q)=\lim_{m\rightarrow \infty}\frac{1}{m\tau}\ln \left\langle e^{q t j(t)} \right \rangle_d  .
\end{equation}
where $m=t/\tau$. The average can be written as
\begin{equation}\label{traj_markov}
\left\langle e^{q t j(t)} \right\rangle_d\!\!=\!\!\sum_{i_0\dots i_{m-1}}\!\!\!\!  A_{i_{m-1} i_{m-2}} e^{q j_{i
_{m-1} i_{m-2}}}\dots  A_{i_{1} i_{0}} e^{q j_{i_{1} i_{0}}}P^{(st)}_{i_0} .
\end{equation}
Here $P^{(st)}_{l}$ is the stationary probability and summation is performed over all the possible trajectories.
We define the column vector $\vec{P}^{(st)}$ having $P_l^{st}$ as components.  Equation \eqref{eq:psid_app} then becomes
\begin{equation}
\psi_d(q)=\lim_{m\rightarrow \infty}\frac{1}{m\tau}\ln \left( \cev{1} \hat{B}^m(q) \vec{P}^{(st)} \right)
\end{equation}
where $\hat{B}$ is defined in Eq.\eqref{eq:tilted_matrix} and $\cev{1}$ is the row vector having all components equal to one. Note that  $\hat{B}$ is a positive matrix and therefore satisfies the Perron-Frobenius theorem. We thus have that $ \cev{1} \hat{B}^m(q) \vec{P}^{(st)} \sim \lambda(q)^m$, where $\lambda(q)$ is the dominant eigenvalue of $\hat{B}$. Performing now the limit $m\rightarrow\infty$ directly yields Eq. \eqref{psid}.

Let us now move to the continuous case. At the first order, the master equation can be written as
\begin{equation}
\vec{P}(t+dt)=\left(\hat{\mathbb{I}}+ \hat{W}dt\right) \vec{P}(t).
\end{equation}
Proceeding as in Eq. \eqref{traj_markov} and substituting the expression for $\hat{W}$ given
by Eq. \eqref{eq:mapping}, the generating function can be expressed in this case as
\begin{eqnarray}
  \left\langle e^{q t j(t)} \right\rangle_c &=& \lim_{dt\rightarrow 0}\cev{1} \left[\hat{\mathbb{I}} + \frac{dt}{\tau} \left(\hat{B}(q)-\hat{\mathbb{I}}\right) \right]^{t/dt}\vec{P}^{(st)}=\nonumber\\
&=&\cev{1}\exp\left[\left(\hat{B}(q)-\hat{\mathbb{I}}\right) \frac{t}{\tau}\right]\vec{P}^{(st)} 
\end{eqnarray}
Substituting this expression in the definition of $\psi_c(q)$ and taking the limit $t\rightarrow \infty$ directly leads to Eq. \eqref{psic}.

\bibliography{uncertainty}

\end{document}